# Security of Public Key Cryptosystems based on Chebyshev Polynomials

Pina Bergamo, Paolo D'Arco, Alfredo De Santis and Ljupco Kocarev *

February 1, 2008

**Abstract**

Chebyshev polynomials have been recently proposed for designing public-key systems. Indeed, they enjoy some nice chaotic properties, which seem to be suitable for use in Cryptography. Moreover, they satisfy a semi-group property, which makes possible implementing a trapdoor mechanism. In this paper we study a public key cryptosystem based on such polynomials, which provides both encryption and digital signature. The cryptosystem works on real numbers and is quite efficient. Unfortunately, from our analysis it comes up that it is not secure. We describe an attack which permits to recover the corresponding plaintext from a given ciphertext. The same attack can be applied to produce forgeries if the cryptosystem is used for signing messages. Then, we point out that also other primitives, a Diffie-Hellman like key agreement scheme and an authentication scheme, designed along the same lines of the cryptosystem, are not secure due to the aforementioned attack. We close the paper by discussing the issues and the possibilities of constructing public key cryptosystems on real numbers.

## 1 Introduction

CHAOS AND CRYPTOGRAPHY. The study of chaotic systems and their possible applications to Cryptography has received considerable attention during the last years in a part of the scientific community. Chaotic systems are indeed characterized by sensitive dependence on initial conditions and similarity to random behavior, properties which seem pretty much the same required by several cryptographic primitives (see [20] for a brief overview).

In [17] for the first time a symmetric key cryptosystem based on Chaos Theory was presented in a well-established cryptographic conference, but it was cryptoanalysed in the same conference [5]. Another scheme based on chaotic maps was broken in [3].

Since then Chaos Theory has not received much attention inside the cryptographic community. However, it has had several applications in other communication areas and people involved in Chaos Theory have been keeping working on the idea of using the properties of chaotic systems in designing efficient cryptographic primitives.

Two main approaches to the use of chaotic systems in designing cryptographic systems can be found in the literature. One of these approaches uses hardware-based synchronized chaotic circuits [26] where, in order to encrypt messages, the cleartext is hidden in the spectral domain of the chaotic signal. This method is strongly related to the concept of synchronization of two chaotic systems and the interested reader can find a survey on the state of art in this field in [19].

*P. Bergamo, P. D'Arco, A. De Santis are with the Dipartimento di Informatica ed Applicazioni, Università degli Studi di Salerno, 84081, Baronissi (SA), Italy. L. Kocarev is with the Institute for Nonlinear Science, University of California San Diego, 9500 Gilman Drive, La Jolla, CA 92093-0402, USA. E-mails: {bergamo, paodar, ads}@dia.unisa.it, lkocarev@ucsd.edu

The other, still for encryption purposes, has investigated the simulation of chaotic discrete dynamical systems on a computer (see [12, 23, 35] to name few).

Many chaotic systems are defined over real numbers. On the other hand, Cryptography deals with systems defined mostly on finite fields. This yields some immediate consequences. Some ordinary design strategies and standard cryptanalytic methods cannot be applied to cryptosystems based on chaotic systems working over real numbers. Just to exemplify, cryptographic systems have secret parameters taking values over a large but finite field. Hence, a brute force attack, which simply tries all elements of the field in searching the secret values might be infeasible but possible. If the range of the parameters of a cryptosystem based on real numbers is a continous infinite interval, an exaustive search is just impossible.

However, at the state of current knowledge, the security of chaos-based cryptosystems defined over real numbers is not well understood.

PUBLIC KEY CRYPTOGRAPHY. Public Key Cryptography enables users who do not share any secret key to securely communicate over a public channel. More precisely, in a public key cryptosystem every user $U$ has a pair of keys $(p_U, s_U)$. The key $p_U$ of user $U$ is public and can be used by everybody else to send an encrypted message to $U$. The key $s_U$ enables to decrypt messages encrypted with key $p_U$, and is kept secret by $U$. Hence, $U$ is the only user able to decrypt encrypted messages. Roughly speaking, the security of a public key cryptosystem, is based on the assumption that computing the secret key $s_U$ given the public key one $p_U$ (even if theoretically possible) is computationally infeasible.

From an historical point of view, Diffie and Hellman, with the publication in 1976 of their paper, *New Directions in Cryptography* [13], introduced the idea[1] of public key cryptography. Later on, Rivest, Shamir and Adlemann, proposed the well-known RSA cryptosystem [30], which realized such an idea. Since then many new cryptosystems have been proposed (see [28, 25] for some relevant examples) and, in general, public key cryptography is a well-established and sound field of knowledge.

THE ISSUE OF SECURITY. Two of the top concerns cryptographers have been dealing with since the idea of public key cryptography was introduced are *what a secure public key cryptosystem is* and *how an efficient one can be constructed*. The first received an answer by Goldwasser and Micali in [16], where the notion of semantic security (w.r.t passive attacks) was established, and by Rackoff and Simon [29], where adaptively chosen ciphertext attacks were considered. The adversary in the latter powerful setting has access to the decryption algorithm and can obtain the plaintexts corresponding to ciphertext messages of his own choosing (apart the challenge ciphertext he has to attack).

However, it turned out to be a difficult task to get an efficient cryptosystem, secure against adaptive chosen ciphertext attacks. Several proposals were given along the years. In 1998 Cramer and Shoup [10] gave the first practical and secure public key encryption scheme. We refer the interested reader to the journal version [11] of such a paper [10] for details about the cryptosystem and for a brief hystorical excursus.

STANDARD MODEL AND RANDOM ORACLE MODEL. The methodology usually applied in Cryptography in order to show that a given protocol meets certain security requirements is *reductionist*: assuming that for a well-known computational problem there are no efficient (i.e., probabilistic polynomial time) algorithms, it is shown that an efficient algorithm breaking the security requirements of the protocol can be used as a building block for constructing an efficient algorithm for solving the supposed to be hard computational problem. In other words, the security of the protocol is reduced to the presumed difficulty of a certain computational problem.

Several currently available public key cryptosystems are defined over finite fields and use modular arithmetics. Their security is often based on the presumed difficulty of solving certain number theoretic problems, like factoring large composite integers, computing the discrete logarithm in finite multiplicative groups, deciding quadratic residuosity of an element, computing square roots and so on. More precisely, two kinds of proofs of security have been given. The first one are proofs in the so called *standard model*, where the security of the scheme is based on standard assumptions, like the aforementioned ones.

---

[1] Even if recently it has been found out that at the GCHQ [18] the idea of public key had already been proposed by the time Diffie and Hellman published their paper but kept secret due to military reasons, it is undoubtly that [13] introduced such an idea into the scientific community

The second deal with the so called *random oracle* model [2], where a sort of idealized model is considered, and the scheme is proved secure in such a model. Then, it is argued that if in the real world the idealized component (i.e., a random function) is opportunely instanciated (e.g., by means of a concrete hash function) the scheme is secure. However, it has been shown that there are schemes secure in the random oracle model but insecure in any implementation in the real world [6]. Hence, the latter are also considered as "heuristic proofs" of security.

EMPIRICAL ANALYSIS. Several well-known and widely used cryptosystems have not been proven secure according to the reductionistic methodology. Formal proofs in the standard model or in the random oracle model have not always been found. Such cryptosystems have been considered secure when they have been used for a long time and no easy method for breaking them has been discovered.

PUBLIC-KEY SCHEMES AS DYNAMICAL SYSTEM. Since 1976, numerous public-key algorithms have been proposed; three most widely used public-key crypto-systems are: RSA, Rabin and ElGamal. From a dynamical point of view, all three encryption algorithms, RSA, ElGamal, and Rabin, employ one single system:

$$X_{n+1} = (X_n)^p \pmod{N}, \tag{1}$$

where $X_n$ is an integer, $0 \le X_n \le N-1$, and, $X_0$, $p$ and $N$ are properly chosen integers. For example, in the ElGamal public-key scheme, one uses (1), where $N$ is a prime, $X_0$ is a generator of the multiplicative group $\mathbb{Z}_N^*$ of integers modulo $N$, and $1 \le p \le N-2$. In the RSA algorithm, $N = PQ$, where $P$ and $Q$ are two random distinct primes, $p$ is an integer $1 < p < \phi$, where $\phi = (P-1)(Q-1)$, such that $\gcd(p, \phi) = 1$, and $X_0$ is the message to be encrypted. Rabin public-key encryption scheme uses (1) with $p = 2$, $N = PQ$, where $P$ and $Q$ primes both congruent to $3 \pmod 4$, and $X_0$ is the message to be encrypted. All three schemes use the the following property of (1):

$$(X^p)^q = X^{pq} \pmod{N}. \tag{2}$$

Recently, several authors have suggested public-key encryption algorithms based on chaotic dynamical systems, defined on real numbers, for which the property (2) is satisfied. Since in this paper we only consider dynamical systems defined over real numbers and enjoying property (2), we refer the reader to Section 2 of [24] for a brief overview on some previously proposed public key cryptosystems based on different chaotic systems and for some more references to the subject.

K. Umeno was probably the first author who suggested that a rational map defined by the elliptic function, which can be expressed directly by a rational polynomial [31], can be used in the public-key scenario [32]. In [21] the authors proposed a public-key encryption algorithm and a signature algorithm, using chaotic Chebyshev polynomials, and suggested an alternative implementation by means of some generalised Chebyshev maps (see [31] and [22]), termed Jacobian Elliptic Chebyshev Rational Maps in [22].

OUR CONTRIBUTION. We start by analysing the public-key cryptosystem based on Chaos Theory, described in [21], which uses Chebyshev polynomials. We show that such a cryptosystem, even if efficient and practical, unfortunately, is not secure. Indeed, we describe an attack that permits to recover the corresponding plaintext from a given ciphertext. The same attack can be applied to produce forgeries if the cryptosystem is used for signing messages. We also consider a realization of the cryptosystem on the Jacobian Elliptic Chebyshev Rational Maps. We show that the attack works against this cryptosystem as well. Then, we point out that also other primitives, a Diffie-Hellman like key agreement scheme [32] and an authentication scheme [37], designed along the same lines of the cryptosystem, are not secure due to the aforementioned attack. We close the paper by discussing the main issues concerning with the design and the implementation of public key systems that work on real numbers, summarising our results, and outlining some possible research directions.

## 2 Chebyshev Polynomials

In this section we briefly describe Chebyshev polynomials, since they represent the cornerstone on which the public key cryptosystem, described in [21], and the authentication scheme, described in [37], are built.

**Definition 2.1** *Let $n$ be an integer, and let $x$ be a variable taking value over the interval $[-1, 1]$. The polynomial $T_n(x) : [-1, 1] \to [-1, 1]$ is recursively defined as*

$$T_n(x) = 2 \cdot x \cdot T_{n-1}(x) - T_{n-2}(x), \quad \text{for any } n \geq 2,$$

where $T_0(x) = 1$ and $T_1(x) = x$.

Some examples of Chebyshev polynomials are (see Fig. 1):

$T_2(x) = 2 \cdot x^2 - 1$
$T_3(x) = 4 \cdot x^3 - 3 \cdot x$
$T_4(x) = 8 \cdot x^4 - 8 \cdot x^2 + 1$

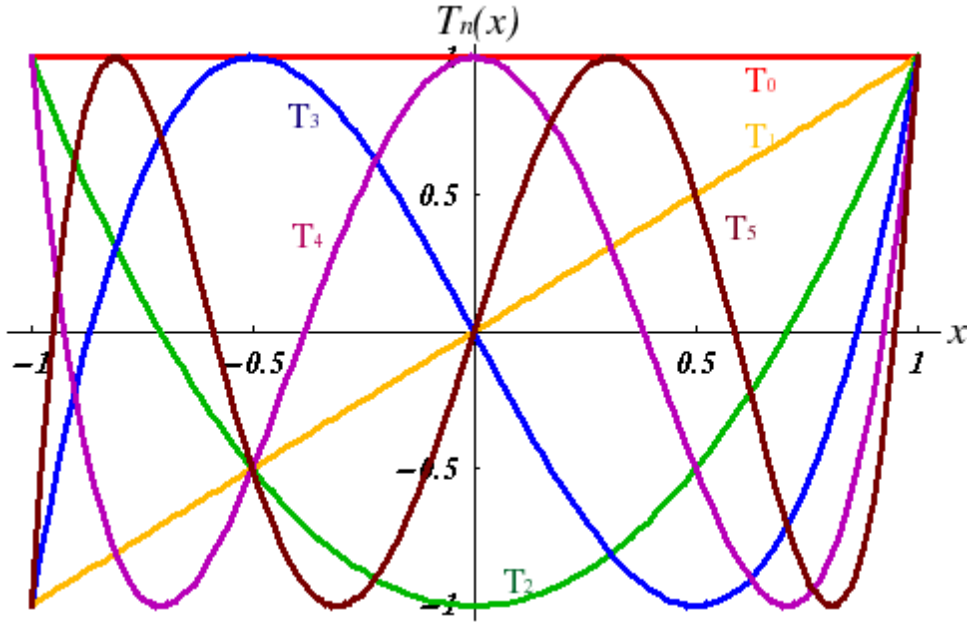

Figure 1: Chebyshev polynomials

One of the most important properties of Chebyshev polynomials is the so called *semi-group property* which establishes that:

$$T_r(T_s(x)) = T_{r \cdot s}(x). \tag{3}$$

An immediate consequence of this property is that Chebyshev polynomials *commute under composition*:

$$T_r(T_s(x)) = T_s(T_r(x)).$$

# 3 A Cryptosystem based on Chebyshev Polynomials

A public key cryptosystem based on Chebyshev polynomials was proposed in [21]. It can be viewed as a generalization of the ElGamal public-key cryptosystem [14].

## 3.1 The Cryptosystem

The cryptosystem is composed of three algorithms: a Key Generation algorithm, an Encryption algorithm, and a Decryption algorithm.

**Key Generation Algorithm.** Key Generation takes place in three steps.

> *Alice*, in order to generate the keys, does the following:
>
> 1. Generates a large integer $s$.
>
> 2. Selects a random number $x \in [-1, 1]$ and computes $T_s(x)$.
>
> 3. *Alice* sets her public key to $(x, T_s(x))$ and her private key to $s$.

**Encryption Algorithm.** Encryption requires five steps:

> *Bob*, in order to encrypt a message, does the following:
>
> 1. Obtains *Alice*'s authentic public key $(x, T_s(x))$.
>
> 2. Represents the message as a number $M \in [-1\ 1]$.
>
> 3. Generates a large integer $r$.
>
> 4. Computes $T_r(x), T_{r \cdot s}(x) = T_r(T_s(x))$ and $X = M \cdot T_{r \cdot s}(x)$.
>
> 5. Sends the ciphertext $C = (T_r(x), X)$ to *Alice*.

**Decryption algorithm.** Decryption requires two steps:

> *Alice*, to recover the plaintext $M$ from the ciphertext $C$, does the following:
>
> 1. Uses her private key $s$ to compute $T_{s \cdot r}(x) = T_s(T_r(x))$.
>
> 2. Recovers $M$ by computing $M = X/T_{s \cdot r}(x)$.

## 3.2 Correctness of the Cryptosystem

The algorithm is correct due to the semi-group property of the Chebyshev polynomials. Indeed, encryption provides:

$$X = M \cdot T_r(T_s(x))$$

Since Chebyshev polynomials commute under composition, it follows that:

$$X = M \cdot T_s(T_r(x))$$

Therefore:

$$M = X/T_{s \cdot r}(x)$$

## 3.3 Implementation

Both encryption and decryption involve the evaluation of Chebyshev polynomials. If we evaluate Chebyshev polynomials directly, applying the recursive definition, then the computation of $T_n(x)$ takes linear time in $n$. However, it is possible to further reduce the computation to a logarithmic number of steps [15], by noticing that

$$\begin{aligned} T_{2 \cdot n}(x) &= T_2(T_n(x)), \\ T_{2 \cdot n+1}(x) &= 2 \cdot T_{n+1}(x) \cdot T_n(x) - x, \end{aligned}$$

and re-organizing the computation. More precisely, we can use the recursive relation for evaluating Chebyshev polynomials:

$$\begin{array}{l} T_0 = 1 \\ T_1 = x \\ T_n(x) = \left\{ \begin{array}{ll} 2 \cdot T_{n/2}^2(x) - 1 & \text{if } n \text{ is even} \\ 2 \cdot T_{(n-1)/2}(x) \cdot T_{(n+1)/2}(x) - x & \text{otherwise} \end{array} \right. \end{array}$$

Another important issue that must be considered when implementing the above cryptosystem is the finite precision of the arithmetics. In [21] the authors pointed out that the semi-group property of Chebyshev polynomials, stated by equation (3), holds only if the values $s$ and $r$, chosen by Alice and Bob, are such that $s < s_0$ and $r < r_0$, where $s_0$ and $r_0$ are constant values depending on the arithmetics precision used in implementing the encryption and decryption algorithms. They gave a table where, for certain precisions, expressed in terms of bits, some possible upper bound for $s_0$ and $r_0$ hold. For example, a 2048-bit precision implies constants $s_0$ and $r_0$ smaller than $2^{970}$. Such upper bounds where empirical determined. No general relation linking the arithmetic precision of the operations to the values of $s_0$ and $r_0$ is currently known.

## 4 Security Analysis of the Cryptosystem

In this section we show that the above cryptosystem is not secure. Given a ciphertext an adversary, by exploiting the same definition of Chebyshev polynomials and after some algebra, can recover the cleartext.

In [21] it was presumed to be secure based on the following observation: as pointed out the scheme resembles ElGamal encryption scheme. The security of ElGamal encryption scheme is based on the intractability of the discrete logarithm problem in $Z_n^*$, i.e., given $n$, $x$ and $x^p$, find $p$. In the above scheme, given $x$ and $T_p(x)$, the value $T_p(x)$ is the value of a polynomial of order $p$, *not just* a power $x^p$. Hence, computing the order of the polynomial $p$, given only one pair $(x, T_p(x))$ seems to be much harder than computing $p$ from a power. Thus, recovering $s$ given $x$ and $T_s(x)$ seems only possible by computing $T_p(x)$ for all $p > 2$ and, then, comparing for which $p$ the equality $T_p(x) = T_s(x)$ holds.

Unfortunately, there are some fundamental differences between the two schemes: ElGamal scheme is implemented over $Z_n^*$ and uses modular arithmetic. Then, given $x$ and $x^p$ the discrete logarithm is uniquely determined while, as we will show later, there are several Chebyshev polynomials passing through the same point.

### 4.1 How to Recover the Plaintext

In this section we present an attack which enables an adversary to recover from a given ciphertext the corresponding cleartext.

First of all, we will use the trigonometric functions $\cos(x)$ and $\arccos(x)$ defined as

$$\cos : R \to [-1,1] \quad \text{and} \quad \arccos : [-1,1] \to [0,\pi].$$

The $\cos(x)$ function has period $2\pi$.

Notice that Chebyshev polynomials can be alternatively defined as follows:

**Definition 4.1** *Let $n$ be an integer, and let $x$ be a variable taking value over the interval $[-1,1]$. The polynomial $T_n(x) : [-1,1] \to [-1,1]$ is defined as:*

$$T_n(x) = \cos(n \cdot \arccos(x)).$$

A simple trigonometric argument shows that Definition 2.1 and Definition 4.1 are equivalent.

**Description of the Attack.** Let $(x, T_s(x))$ be Alice's public key. In order to encrypt a message $M$, Bob chooses a large integer r and computes:

$$T_r(x), \quad T_{r \cdot s}(x) = T_r(T_s(x)), \quad \text{and} \quad X = M \cdot T_{r \cdot s}(x)$$

Then, he sends the cipher-text $C = (T_r(x), X)$ to Alice.

Unfortunately an adversary, given Alice's public key $(x, T_s(x))$ and the ciphertext $(T_r(x), X)$, can recover $M$ as follows:

> The adversary, to get the message, does the following:
> 
> 1. Computes an $r'$ such that $T_{r'}(x) = T_r(x)$.
> 2. Evaluates $T_{r's}(x) = T_{r'}(T_s(x))$.
> 3. Recovers $M = \frac{X}{T_{r's}(x)}$.

The attack is always successful because, if $r'$ is such that $T_{r'}(x) = T_r(x)$, then:

$$\begin{aligned}
T_{r \cdot s}(x) &= T_{s \cdot r}(x) \\
&= T_s(T_r(x)) = T_s(T_{r'}(x)) \\
&= T_{s \cdot r'}(x) = T_{r' \cdot s}(x) \\
&= T_{r'}(T_s(x)).
\end{aligned}$$

Let us show how such an $r'$ can be computed. Let $\mathcal{N}$ be the set of natural numbers and let $\mathcal{Z}$ be the set of integers. According to Definition 4.1, it holds that $T_r(x) = \cos(r \cdot \arccos(x))$. Let

$$\mathcal{P} = \left\{ \frac{\pm \arccos(T_r(x)) + 2k\pi}{\arccos(x)} \ \bigg| \ k \in \mathcal{Z} \right\}.$$

Notice that some $r'$ belonging to the set $\mathcal{P}$ might not be integers. However, the following result shows that $\mathcal{P}$ contains *all* possible integers $r'$ defining polynomials $T_{r'}(x)$ passing through $T_r(x)$.

**Lemma 4.2** *For each pair $(x, T_r(x))$, the integer $r'$ satisfies $T_{r'}(x) = T_r(x)$ if and only if $r' \in \mathcal{P} \cap \mathcal{N}$.*

**Proof.** Let $r' \in \mathcal{P} \cap \mathcal{N}$. Assume that

$$r' = \frac{\arccos(T_r(x)) + 2k'\pi}{\arccos(x)}$$

for a certain $k'$. By using Definition 4.1, it holds that

$$\begin{aligned}
T_{r'}(x) &= \cos\left(r' \arccos(x)\right) \\
&= \cos\left(\frac{\arccos(T_r(x)) + 2k'\pi}{\arccos(x)} \cdot \arccos(x)\right) \\
&= \cos\left(\arccos(T_r(x)) + 2k'\pi\right) \\
&= \cos\left(\arccos(T_r(x))\right) \\
&= T_r(x).
\end{aligned}$$

Hence, if $r' \in \mathcal{P} \cap \mathcal{N}$, then $T_{r'}(x) = T_r(x)$. If $r' = \frac{-\arccos(T_r(x)) + 2k'\pi}{\arccos(x)}$ we can apply exactly the same argument.

On the other hand, assume that $T_{r'}(x) = T_r(x)$ for a certain $r' \in \mathcal{N}$. Then,

$$T_{r'}(x) = \cos\left(r' \arccos(x)\right) = T_r(x).$$

Applying the arccos function to both members, we get:

$$\arccos\left(\cos\left(r' \arccos(x)\right)\right) = \arccos(T_r(x)). \tag{4}$$

Let $y = \arccos(w)$. Due to the equality $\cos(-\beta) = \cos(\beta)$, for every angle $\beta$, and due to the periodicity of the cos function, all angles $\beta$ such that $\cos(\beta) = w$ are given by $\beta = \pm y + 2k\pi$, for $k \in \mathcal{Z}$. Therefore, identity (4) holds if and only if

$$r' \arccos(x) = \pm \arccos(T_r(x)) + 2k'\pi.$$

where $k' \in \mathcal{Z}$. Dividing both members by $\arccos(x)$, we get:

$$r' = \frac{\pm \arccos(T_r(x)) + 2k'\pi}{\arccos(x)}$$

i.e., $r' \in \mathcal{P} \cap \mathcal{N}$. Thus, the lemma holds. □

Using the above result, denoting by

$$a = \frac{\arccos(T_r(x))}{\arccos(x)} \quad \text{and} \quad b = \frac{2\pi}{\arccos(x)} \tag{5}$$

the adversary has to find an integer $k \in \mathcal{Z}$ and a positive integer $u \in \mathcal{N}$ solutions to one of the two equations

$$a + k \cdot b = u \quad \text{or} \quad -a + k \cdot b = u \tag{6}$$

given $a$ and $b$.

Let $(a \bmod 1)$ and $(b \bmod 1)$ be the fractional parts of $a$ and $b$. The actual problem becomes solving

$$(a \bmod 1) + k \cdot (b \bmod 1) = z$$

or

$$-(a \bmod 1) + k \cdot (b \bmod 1) = z.$$

**How to find $k$ in a real implementation.** Assume that we use a finite precision implementation in base $B \geq 2$, and that $L$ is the maximum number of digits of $(a \bmod 1)$ and $(b \bmod 1)$. Then, multiplying all terms by $B^L$, we can rewrite the above equations in equivalent form as

$$(a \bmod 1) \cdot B^L + k \cdot (b \bmod 1) \cdot B^L = z \cdot B^L$$

and
$$-(a \bmod 1) \cdot B^L + k \cdot (b \bmod 1) \cdot B^L = z \cdot B^L.$$

Denoting by $a'$ the integer $(a \bmod 1) \cdot B^L$ and by $b'$ the integer $(b \bmod 1) \cdot B^L$, the solutions to the above equations are exactly the solutions to the linear modular equations

$$b' \cdot k \equiv a' \bmod B^L \quad \text{and} \quad b' \cdot k \equiv -a' \bmod B^L. \tag{7}$$

However, notice that we can restrict our attention to just one of the above modular equations. Indeed, since $b' \cdot k \equiv -a' \bmod B^L$ is equivalent to $b' \cdot (-k) \equiv a' \bmod B^L$, once we have solved $b' \cdot k \equiv a' \bmod B^L$, we easily derive the solutions to the second one. More precisely, if $k$ is solution to $b' \cdot k \equiv a' \bmod B^L$ then $B^L - k$ is solution to $b' \cdot k \equiv -a' \bmod B^L$.

We can get efficiently the set of solutions to linear modular equations of the form $b'k \equiv a' \bmod B^L$ (see, for example, Chap. 33 of [9]). Denoting by $<b'> = \{b'j \bmod B^L | j \in Z_{B^L}\}$ the subgroup of elements of $Z_{B^L}^*$ generated by $b'$, it is easy to see that the modular equation has solutions if and only if $a' \in <b'>$. Moreover, denoting by $d$ the $\gcd(b', B^L)$, the above membership condition is equivalent to $d|a'$. The set of distinct solutions to $b'k \equiv a' \bmod B^L$ (if there exist) has cardinality $d$ and is given by

$$x_j = x_0 + j \cdot \frac{B^L}{d} \bmod B^L, \text{ for } j = 1, \ldots, d-1,$$

where the first solution $x_0$ can be obtained directly by applying the Extended Euclidean Algorithm. Indeed, such an algorithm, on input $(b', B^L)$, outputs a triple $(d, s', t')$ of integers where $d = b's' + B^L t'$, and it is easy to check that $x_0 = s' \frac{a'}{d}$ is solution to $b'k \equiv a' \bmod B^L$. From a computational point of view, the above procedure is efficient since the running time of the Extended Euclidean Algorithm requires $O(\log B^L)$ steps in the worst case.

Coming back to our setting, notice that the equations given in (7) have solutions *by construction*. More precisely, there are exactly $d = \gcd(b', B^L)$ distint solutions for each of them, which can be easily found applying the above method. Clearly, just one solution suffices to the adversary's goal.

## 4.2 An Example

We show how an adversary, given Alice's public key $(x, T_s(x))$ and the ciphertext $C = (T_r(x), X)$, where $X = M \cdot T_{rs}(x)$, constructed by Bob in order to send $M$ to Alice, computes the value $T_{rs}(x)$. Then, dividing $X$ by $T_{rs}(x)$, he recovers $M$.

Let us start by generating Alice's public key parameters.

Let $B = 10$, $\pi = 3.141592654$, $x = 0.64278761$ and $s = 106000$. Then, $\arccos(x) = \frac{5}{18}\pi$, and $T_s(x) = \cos(s \cdot \arccos x) = \cos(106000 \cdot \frac{5}{18}\pi) = 0.173648178$. Hence, Alice's public key is given by the pair

$$(x, T_s(x)) = (0.64278761, 0.173648178).$$

Assume that Bob, in order to encrypt a message M, chooses $r = 81500$. Then,

$$T_r(x) = \cos(r \cdot \arccos x) = \cos(81500 \cdot \frac{5}{15}\pi) = -0.939692621,$$

and

$$T_r(T_s(x)) = \cos(r \cdot \arccos(T_s(x))) = \cos(81500 \cdot \frac{4}{9}\pi) = 0.766044443.$$

By applying the strategy described before, an adversary computes an $r'$ such that $T_{r'}(T_s(x)) = T_r(T_s(x))$. Since it holds that

$$\arccos(T_r(x)) = \frac{8\pi}{9} \text{ and } \arccos(x) = \frac{5\pi}{18},$$

the set of possible integer indices $r'$ is given by

$$\mathcal{P} = \left\{ \frac{\pm \arccos(T_r(x))}{\arccos(x)} + \frac{2\pi k}{\arccos(x)} \;\bigg|\; k \in \mathcal{Z} \right\} = \{\pm 3.2 + 7.2k \;|\; k \in \mathcal{Z}\}.$$

Hence, the adversary has to find a solution to one of the follwoing two equations

$$3.2 + 7.2k_1 = u_1 \text{ and } -3.2 + 7.2k_2 = u_2. \tag{8}$$

where $u_1, u_2 \in \mathcal{N}$. By considering only the fractional parts, the problem becomes solving one of

$$0.2 + 0.2k_1 = z_1, \text{ or } -0.2 + 0.2k_2 = z_2$$

where $z_1, z_2 \in \mathcal{N}$. Since $L = 1$, then $B^L = 10$, and the above equations are equivalent to

$$2 + 2k_1 = 10z_1 \text{ and } -2 + 2k_2 = 10z_2$$

whose solutions are exactly the solutions to the modular equations

$$2k \equiv 8 \bmod 10 \quad \text{and} \quad 2k \equiv 2 \bmod 10. \tag{9}$$

Let us consider the first one. This equation has solutions since $\gcd(2, 10) = 2$ and $2|8$. Precisely, there are 2 solutions, given by $k = 4 + i5$, for $i = 0, 1$, where 4 is the solution $x_0$ obtained directly by means of the Extended Euclidean Algorithm. By choosing one of them, for example 4, the corresponding index $r'$, computed evaluating the first one of (8) is 32. Then, it holds that:

$$T_{32}(T_s(x)) = \cos(32 \cdot \frac{4}{9}\pi) = 0.766044443.$$

Hence, the adversary has computed $T_{rs}(x)$. The cleartext sent by Bob is computed by the adversary as $X/T_{32s}(x)$.

For completeness, notice the two solutions to the second equation are $\{6, 1\}$ and are obtained by computing $-4 \bmod 10$ and $-9 \bmod 10$. By choosing one of them, for example 1, the corresponding index $r'$, computed evaluating the second of (8) is 4. Then, it holds that:

$$T_4(T_s(x)) = \cos(4 \cdot \frac{4}{9}\pi) = 0.766044443.$$

Hence, the adversary has computed $T_{rs}(x)$. The cleartext sent by Bob is computed by the adversary as $X/T_{4s}(x)$.

## 5 A Cryptosystem based on Jacobian Elliptic Chebyshev Rational Maps

As suggested in [21], instead of using Chebyshev polynomials, the cryptosystem we have previously analised can be also realized by using the *Jacobian Elliptic Chebyshev Rational Maps*, studied in [31] and [22]. In the following subsections we show how to implement such a cryptosystem. Then, we show that the attack we have identified for the cryptosystem based on Chebyshev polynomials applies to this cryptosystem as well.

## 5.1 Jacobian Elliptic Chebyshev Rational Maps

The Jacobian Elliptic Chebyshev Rational Maps are rational functions defined as follows [22]:

**Definition 5.1** *Let $p$ be a positive integer, let $\omega \in [-1, 1]$ be a real number, and let $k \in [0, 1]$ be a real number called modulus. Jacobian Elliptic Chebyshev Rational Maps are defined by*

$$R_{p+1}(\omega, k) = \frac{2\omega}{1 - k^2(1 - R_p(\omega, k)^2)(1 - \omega^2)} R_p(\omega, k) - R_{p-1}(\omega, k),$$

*where $R_0(\omega, k) = 1$ and $R_1(\omega, k) = \omega$.*

Notice that, when the modulus $k = 0$, the Jacobian Elliptic Chebyshev Rational Map $R_p(\omega, 0)$ is exactly a Chebyshev polynomial, i.e., $R_p(\omega, 0) = T_p(\omega)$.

Jacobian Elliptic Chebyshev Rational Maps enjoy the *semi-group property*. Indeed, for each integers $r, s \geq 2$, and for each $\omega, k$, it holds that

$$R_r(R_s(\omega, k), k) = R_{r \cdot s}(\omega, k). \tag{10}$$

Hence, these maps commute under composition, i.e.,

$$R_r(R_s(\omega, k), k) = R_s(R_r(\omega, k), k).$$

## 5.2 The Cryptosystem

The cryptosystem is composed of three algorithms: a Key Generation algorithm, an Encryption algorithm, and a Decryption algorithm.

**Key Generation Algorithm.** Key Generation takes place in three steps:

> *Alice*, in order to generate the keys, does the following:
>
> 1. Generates a large integer $s$.
>
> 2. Selects two random numbers $\omega \in [-1, 1]$ and $k \in [0, 1]$, and computes $R_s(\omega, k)$.
>
> 3. *Alice* sets her public key to $(\omega, k, R_s(\omega, k))$ and her private key to $s$.

**Encryption Algorithm.** Encryption requires five steps:

> *Bob*, in order to encrypt a message, does the following:
>
> 1. Obtains *Alice*'s authentic public key $(\omega, k, R_s(\omega, k))$.
>
> 2. Represents the message as a number $M \in [-1, 1]$.
>
> 3. Generates a large integer $r$.
>
> 4. Computes $R_r(\omega, k), R_{r \cdot s}(\omega, k) = R_r(R_s(\omega, k), k)$, and $X = M \cdot R_{r \cdot s}(\omega, k)$.
>
> 5. Sends the ciphertext $C = (R_r(\omega, k), X)$ to *Alice*.

**Decryption Algorithm.** Decryption requires two steps:

> *Alice*, to recover the plaintext $M$ from the ciphertext $C$, does the following:
>
> 1. Uses her private key $s$ to compute $R_{s \cdot r}(\omega, k) = R_s(R_r(\omega, k), k)$.
>
> 2. Recovers $M$ by computing $M = X / R_{s \cdot r}(\omega, k)$.

Notice that, the value of $k$, which defines the form of the map, could be the same for all users of the system.

## 5.3 Correctness of the Cryptosystem

The cryptosystem is correct due to the semi-group property of the Jacobian Elliptic Chebyshev Rational Maps. Indeed, encryption provides:

$$X = M \cdot R_r(R_s(\omega, k), k)$$

Since the maps commute under composition, it follows that:

$$X = M \cdot R_s(R_r(\omega, k), k).$$

Therefore:

$$M = X / R_{s \cdot r}(\omega, k).$$

## 5.4 Jacobian Elliptic Functions and Jacobian Elliptic Chebyshev Rational Maps

Jacobian elliptic Chebyshev rational maps can be equivalently defined by means of the Jacobian elliptic functions [22].

Let $\omega \in [-1, 1]$, let $k \in [0, 1]$, and let $\varphi \in [0, 2\pi]$ be the angle, referred to as the *amplitude* of $\omega$, defined by

$$\omega = \int_0^\varphi \frac{d\theta}{(1 - k^2 \cdot \sin^2(\theta))^{\frac{1}{2}}}.$$

Then, the Jacobian elliptic functions $sn(\omega, k)$ and $cn(\omega, k)$ are defined as follows:

$$sn(\omega, k) = \sin(\varphi) \quad \text{and} \quad cn(\omega, k) = \cos(\varphi).$$

Let $k' = \sqrt{1 - k^2}$. The above functions are doubly-periodic, having a real period and an imaginary one. More precisely, denoting by

$$K = \int_0^{\frac{\pi}{2}} \frac{d\theta}{(1 - k^2 \cdot \sin^2(\theta))^{\frac{1}{2}}} \quad \text{and} \quad iK' = i \int_0^{\frac{\pi}{2}} \frac{d\theta}{(1 - k'^2 \cdot \sin^2(\theta))^{\frac{1}{2}}}$$

where $i$ is the imaginary unit, we get that $sn(\omega, k)$ has periods $4K$ and $2iK'$; while $cn(\omega, k)$ has periods $4K$ and $2K + 2iK'$. We restrict our attention to the real periodicity.

For any fixed $k$, the function $cn^{-1}(v, k)$, inverse of the Jacobian elliptic function $cn(\omega, k)$, relatively to the interval $[0, 2K]$, is given by:

$$cn^{-1}(v, k) = \int_0^\varphi \frac{d\theta}{(1 - k^2 \sin^2(\theta))^{\frac{1}{2}}}.$$

where $\varphi = \arccos(v)$.

Then, we can state the following alternative definition for the Jacobian elliptic Chebyshev rational maps:

**Definition 5.2** *Let $p \geq 2$ be an integer, let $k \in [0,1]$ be a real number, and let $\omega \in [-1,1]$. The Jacobian elliptic Chebyshev rational maps with modulus $k$ are defined by*

$$R_p(\omega, k) = cn(p \cdot cn^{-1}(\omega, k), k).$$

## 5.5 Efficient Computation of $cn(\omega, k)$, $sn(\omega, k)$, and $cn^{-1}(v, k)$,

The functions $cn(\omega, k)$, $sn(\omega, k)$, and $cn^{-1}(v, k)$, all defined in terms of elliptic integrals, can be efficiently computed by means of the *Arithmetic-Geometric Method*, (A.G.M. method, for short). Roughly speaking, such a method works as follows: starting with $(a_0, b_0)$, it proceeds to determine number triples

$$(a_1, b_1, c_1), (a_2, b_2, c_2), \ldots, (a_n, b_n, c_n)$$

according to the following scheme of arithmetic and geometric mean:

$$a_{j+1} = \frac{1}{2}(a_j + b_j) \quad b_{j+1} = (a_j \cdot b_j)^{\frac{1}{2}} \quad \text{and} \quad c_{j+1} = \frac{1}{2}(a_j - b_j).$$

Assume that we use an arithmetic in base $B$ with $N$-digit precision of the operations. The procedure stops at the $n$-th step when $a_n = b_n$, i.e., when $c_n = 0$. Notice that such an equality is achieved when the relative error $\epsilon_n = 1 - \frac{b_n}{a_n}$ is less than the degree of accurancy fixed by the implementation i.e., $B^{-N}$. It has been estimated (see, for example [33]) that the relative error $\epsilon_j = 1 - \frac{b_j}{a_j}$ decays approximatively as $\epsilon_j \approx \frac{1}{8}e^{-2^j}$, from which it easily follows that the method converges after roughly $\log N$ steps.

To compute the functions $cn(\omega, k)$ and $sn(\omega, k)$, we apply the A.G.M method starting with $a_0 = 1$, and $b_0 = k'$. Once the A.G.M method stops, we compute the angle (in degrees) $\phi_n = 2^n a_n \omega \frac{180}{\pi}$. Then, applying, for $j = n, \ldots, 1$, the recurrence relation $\sin(2\phi_{j-1} - \phi_j) = \frac{c_j}{a_j} \sin \phi_j$, we compute the angles $\phi_{n-1}, \phi_{n-2}, \ldots, \phi_0$. Finally,

$$sn(\omega, k) = \sin \phi_0 \quad \text{and} \quad cn(\omega, k) = \cos \phi_0.$$

On the other hand, to evaluate $cn^{-1}(v, k)$, for $j = 0, \ldots, n-1$, by applying the recurrence relation $\tan(\gamma_{j+1} - \gamma_j) = \frac{b_j}{a_j} \tan \gamma_j$, where $\gamma_0 = \varphi$, we compute the angles $\gamma_1, \ldots, \gamma_n$, and then

$$cn^{-1}(v, k) = \frac{\gamma_n}{2^n a_n}.$$

Notice that the quarter-period $K$ can be easily computed as well, since it is a special case of the computation of $cn(\omega, k)$ (just set the angle $\varphi = \pi/2$). The reader is referred to [1] for further details on the A.G.M method, and on the computation of $sn(\omega, k)$, $cn(\omega, k)$, and $cn^{-1}(v, k)$. Moreover, an efficient implementation of the above functions can be found in [34].

## 6 Security Analysis of the Cryptosystem

Apart the complexity of the mathematical objects we are dealing with, the attack we have applied against the public key scheme based on Chebyshev polynomials still works against the cryptosystem based on Jacobian elliptic Chebyshev rational maps.

### 6.1 How to Recover the Plaintext

Let $(\omega, k, R_s(\omega, k))$ be Alice's public key. In order to encrypt a message $M$, Bob chooses a large integer $r$ and computes:

$$R_r(\omega, k), \quad R_{r \cdot s}(\omega, k) = R_r(R_s(\omega, k), k), \quad \text{and} \quad X = M \cdot R_{r \cdot s}(\omega, k)$$

Then, he sends the ciphertext $C = (R_r(\omega, k), X)$ to Alice.

Unfortunately an adversary, given Alice's public key $(\omega, k, R_s(\omega, k))$ and the ciphertext $C = (R_r(\omega, k), X)$, can recover $M$ as follows:

> The adversary, to get the message, does the following:
> 1. Computes an $r'$ such that $R_{r'}(\omega, k) = R_r(\omega, k)$.
> 2. Evaluates $R_{r's}(\omega, k) = R_{r'}(R_s(\omega, k), k)$.
> 3. Recovers $M = \frac{X}{R_{r's}(\omega, k)}$.

The attack is always successful because, if $r'$ is such that $R_{r'}(\omega, k) = R_r(\omega, k)$, then:

$$R_{r \cdot s}(\omega, k) = R_{s \cdot r}(\omega, k) = R_s(R_r(\omega, k), k) = R_s(R_{r'}(\omega, k), k)$$
$$= R_{s \cdot r'}(\omega, k) = R_{r' \cdot s}(\omega, k) = R_{r'}(R_s(\omega, k), k).$$

Let us show how such an $r'$ can be computed. According to Definition 5.2, it holds

$$R_r(\omega, k) = cn(r \cdot cn^{-1}(\omega, k), k).$$

Hence, applying the $cn^{-1}$ function to both members of the equality, and using the periodicity of $cn(\omega, k)$ and the property $cn(\omega, k) = cn(-\omega, k)$, we get that

$$\pm cn^{-1}((R_r(\omega, k), k) + z \cdot 4K = r \cdot cn^{-1}(\omega, k),$$

for $z \in \mathcal{Z}$. Notice that we are only considering the real periodicity, since we are not interested in imaginary solutions. Let

$$\mathcal{P} = \left\{ \frac{\pm cn^{-1}(R_r(\omega, k), k) + z \cdot 4K}{cn^{-1}(\omega, k)} \;\middle|\; z \in \mathcal{Z} \right\}$$

We can show that $\mathcal{P}$ contains *all* possible integers $r'$ defining maps $R_{r'}(\omega, k)$ passing through $R_r(\omega, k)$, for certain $r, \omega$, and $k$. The proof proceeds along the same lines of the proof provided for Lemma 1. We omit it since it is essentially the same.

**Lemma 6.1** *For each triple $(\omega, k, R_r(\omega, k))$, the integer $r'$ satisfies $R_{r'}(\omega, k) = R_r(\omega, k)$ if and only if $r' \in \mathcal{P} \cap \mathcal{N}$.*

Setting $a = \frac{cn^{-1}(R_r(\omega, k), k)}{cn^{-1}(\omega, k)}$ and $b = \frac{4K}{cn^{-1}(\omega, k), k)}$ as in (5), we apply *exactly* the same steps we have done in Subsection 4.1 describing the attack against the cryptosystem based on Chebyshev polynomials. Hence, an adversary can recover the plaintext from the ciphertext.

## 6.2 An Example

We show how an adversary, given Alice's public key $(\omega, k, R_s(\omega, k))$ and the ciphertext $C = (R_r(\omega, k), X)$, where $X = M \cdot R_{rs}(\omega, k)$, constructed by Bob in order to send $M$ to Alice, computes the value $R_{rs}(\omega, k)$. Then, dividing $X$ by $R_{rs}(\omega, k)$, he recovers $M$.

Let us start by generating Alice's public key parameters.

Let $B = 10$, $\omega = 0.435946$, $k = 0.3$, and $s = 2342$. Then, $R_s(\omega, k) = cn(s \cdot cn^{-1}(\omega, k), k) = 0.245756$. Hence, Alice's public key is given by the triple

$$(\omega, k, R_s(\omega, k)) = (0.435946, 0.3, 0.245756).$$

Assume that Bob, in order to encrypt a message M, chooses $r = 1876$. Then,

$$R_r(\omega, k) = cn(r \cdot cn^{-1}(\omega, k), k) = -0.938538$$

and

$$R_r(R_s(\omega, k), k) = cn(r \cdot cn^{-1}(R_s(\omega, k), k), k) = 0.613408.$$

By applying the strategy described in Subsection 6.1, an adversary computes an $r'$ such that $R_{r'}(R_s(\omega, k), k) = R_r(R_s(\omega, k), k)$. The set of possible integer indices $r'$ is given by

$$\mathcal{P} = \left\{ \frac{\pm cn^{-1}(R_r(\omega, k), k) + z \cdot 4K}{cn^{-1}(\omega, k)} \mid z \in \mathcal{Z} \right\} = \{\pm 2.6 + 5.8k \mid k \in \mathcal{Z}\}.$$

Hence, the adversary has to find a solution to one of the follwoing two equations

$$2.6 + 5.8k_1 = u_1 \text{ and } -2.6 + 5.8k_2 = u_2. \tag{11}$$

where $u_1, u_2 \in \mathcal{N}$. By considering only the fractional parts, the problem becomes solving one of

$$0.6 + 0.8k_1 = z_1, \text{ or } -0.6 + 0.8k_2 = z_2$$

where $z_1, z_2 \in \mathcal{N}$. Since $L = 1$, then $B^L = 10$, and the above equations are equivalent to

$$6 + 8k_1 = 10z_1 \text{ and } -6 + 8k_2 = 10z_2$$

whose solutions are exactly the solutions to the modular equations

$$8k \equiv 4 \bmod 10 \quad \text{and} \quad 8k \equiv 6 \bmod 10. \tag{12}$$

Let us consider the first one. This equation has solutions since $\gcd(8, 10) = 2$ and $2|4$. Precisely, there are 2 solutions, given by $k = 3 + i5$, for $i = 0, 1$, where 3 is the solution $x_0$ obtained directly by means of the Extended Euclidean Algorithm. By choosing one of them, for example 3, the corresponding index $r'$, computed evaluating the first one of (11) is 20. Then, it holds that:

$$R_{20}(R_s(\omega, k), k) = cn(20 \cdot cn^{-1}(R_s(\omega, k), k), k) = 0.613408.$$

Hence, the adversary has computed $R_{rs}(\omega, k)$. The cleartext sent by Bob is computed by the adversary as $X/R_{20s}(\omega, k)$.

# 7 Key Agreement by using Rational Maps

Rational maps enjoying the semi-group property can be also used to design a Diffie-Hellman like key agreement scheme. Umeno [32] was the first author who suggested such a method.

Let us briefly recall the following definitions, given in [25].

**Definition 7.1** *Key establishment is any process whereby a shared secret key becomes available to two or more parties, for subsequent cryptographic use.*

**Definition 7.2** *A key agreement protocol or mechanism is a key establishment technique in which a shared secret is derived by two or more parties as a function of information contributed by, or associated with, each of these, ideally such that no party can predetermine the resulting value.*

Let us look at the following key agreement protocol:

Let $X$ be a public real value, and let $F(\cdot, \cdot)$ be a rational map enjoying the semi-group property, i.e., $F(p, F(q, X)) = F(pq, X)$.

> *Bob*, in order to agree on a common key with *Alice*, does the following:
>
> 1. Generates a large integer $p$.
> 2. Computes $Y = F(p, X)$.
> 3. Sends $Y$ to Alice.
>
> *Alice*, in order to agree on a common key with *Bob*, does the following:
>
> 1. Generates a large integer $q$.
> 2. Computes $Y' = F(q, X)$.
> 3. Sends $Y'$ to Bob.
>
> Then, *Alice* and *Bob* compute the common value
> $$Z = F(q, Y) = F(q, F(p, X)) = F(p, F(q, X)) = F(p, Y').$$

It is easy to check that if the rational map used in the above scheme is a Chebyshev Polynomial or a Jacobian Elliptic Chebyshev Rational map then, since $X$ is public and $F(p, X)$ and $F(q, X)$ are sent in clear over the channel, an adversary who taps the channel, with no knowledge of the secret values $p$ and $q$, can employ the same attack we have described before for the public-key cryptosystem, and compute the common key.

# 8 Entity Authentication based on Chebyshev Polynomials

Chebyshev Polynomials have also been used to design an authentication scheme. Entity authentication is defined as follows [25]:

**Definition 8.1** *Entity authentication is the process whereby one party is assured (through acquisition of corroborative evidence) of the identity of a second party involved in a protocol, and that the second has actually participated (i.e., is active at, or immediately prior to, the time evidence is acquired).*

In [37] a scheme based on Chebyshev Polynomials, by means of which a user can efficiently authenticate himself to a server in order to log in, was proposed. It strongly resembles the public key cryptosystem described in [21]. Apart minor implementation details, the scheme works as follows:

Let $m \in [-1, 1]$ be a real value, and denote by $T_s^i(\cdot)$ the map $T_s(\cdot)$ iterated $i$ times, i.e., $T_s^i(\cdot) = T_s(T_s(T_s \ldots T_s(\cdot))\ldots) = T_{s^i}(\cdot)$.

> *Setup Phase - Server Side*
>
> 1. The server generates a random number $r$.
> 2. Computes and sends $T_r(m)$ to the user.
>
> *Setup Phase - User Side*
>
> 1. The user chooses a random number $s$.
>
> *i-th Authentication Phase*
>
> 1. The user computes $T_s^i(m)$, and $auth = T_s^i(T_r(m))$, and sends both values to the server.
> 2. The server computes $auth' = T_r(T_s^i(m))$ and checks whether $auth = auth'$. Then, if the check is satisfied, the access is granted.

It is easy to see that, if $m$ and $T_r(m)$ are public, an adversary who gets the messages associated with the first log in request, can apply the same attack we have described before in order to get an integer $s'$ such that $T_{s'}(m) = T_s(m)$. Then, at the i-th session, he can authenticate himself as the real user by computing $T_{s'}^i(m)$, and $auth = T_{s'}^i(T_r(m))$. Indeed, it is easy to show, arguing by induction on $i$, that $T_{s'}^i(m) = T_s^i(m)$. Therefore, it holds that

$$auth = T_{s'}^i(T_r(m)) = T_r(T_{s'}^i(m)) = T_r(T_s^i(m)) = T_s^i(T_r(m)) = auth'.$$

Thus, the scheme is not secure. One way to avoid the above attack is to make $m$ and $T_r(m)$ private to the user and the server. Unfortunately, the scheme is not secure *even if $m$ and $T_r(m)$ are private*. Indeed, even in this scenario, an adversary with no knowledge of the private values $m$ and $T_r(m)$, who just listen to *two consecutive authentication phases*, can subsequently authenticate himself to the server as it were the real user. More precisely, assume that the adversary gets $T_s^{i-1}(m), T_s^{i-1}(T_r(m))$ and $T_s^i(m), T_s^i(T_r(m))$. Then, the attack works as follows:

> The adversary does the following:
>
> 1. Computes an integer $w$ such that $T_w(T_s^{i-1}(m)) = T_s^i(m)$.
> 2. For any $\ell \geq 1$, to authetcate himself at the $(i+\ell)$-th session,
>    (a) Computes
>    $$T_s^{i+\ell}(m) = T_w^\ell(T_s^i(m)) \text{ and } auth = T_s^{i+\ell}(T_r(m)) = T_w^\ell(T_s^i(T_r(m))).$$
>    (b) Sends the pair $(T_s^{i+\ell}(m), auth)$.

Notice that the adversary *does not* need to know the index $i$ of the session. He just needs two consecutive authentication messages.

In order to understand why the attack works, notice that an integer $w$ such that $T_w(T_{s^{i-1}}(m)) = T_{s^i}(m)$ can be computed by applying the same attack we have described before against the cryptosystem. Then, we can proceed by induction on $\ell$ to show that $T_s^{i+\ell}(m) = T_w^\ell(T_s^i(m))$ and $auth = T_s^{i+\ell}(T_r(m)) = T_w^\ell(T_s^i(T_r(m)))$.

Let $\ell = 1$. It is easy to see that

$$T_s^{i+1}(m) = T_s(T_{s^i}(m)) = T_s(T_w(T_{s^{i-1}}(m))) = T_w(T_s(T_{s^{i-1}}(m))) = T_w(T_{s^i}(m)).$$

Then, notice that $T_w(T_{s^{i-1}}(T_r(m))) = T_{s^i}(T_r(m))$. Indeed,

$$T_w(T_{s^{i-1}}(T_r(m))) = T_w(T_r(T_{s^{i-1}}(m))) = T_r(T_w(T_{s^{i-1}}(m))) = T_r(T_{s^i}(m)) = T_{s^i}(T_r(m)).$$

Therefore,

$$T_s^{i+1}(T_r(m)) = T_s(T_{s^i}(T_r(m))) = T_s(T_w(T_{s^{i-1}}(T_r(m)))) = T_w(T_s(T_{s^{i-1}}(T_r(m)))) = T_w(T_{s^i}(T_r(m))).$$

Assume that

$$T_{s^{i+(\ell-1)}}(m) = T_{w^{(\ell-1)}}(T_{s^i}(m)) \text{ and } T_{s^{i+(\ell-1)}}(T_r(m)) = T_{w^{(\ell-1)}}(T_{s^i}(T_r(m))).$$

By applying the inductive hypothesis, it holds that

$$T_{s^{i+\ell}}(m) = T_s(T_{s^{i+(\ell-1)}}(m)) = T_s(T_{w^{(\ell-1)}}(T_{s^i}(m))) = T_s(T_{w^{(\ell-1)}}(T_w(T_{s^{i-1}}(m)))) = T_{w^\ell}(T_{s^i}(m)),$$

and

$$\begin{aligned} T_{s^{i+\ell}}(T_r(m)) = T_s(T_{s^{i+(\ell-1)}}(T_r(m))) &= T_s(T_{w^{(\ell-1)}}(T_{s^i}(T_r(m)))) \\ &= T_s(T_{w^{(\ell-1)}}(T_w(T_{s^{i-1}}(T_r(m))))) \\ &= T_{w^\ell}(T_{s^i}(T_r(m))). \end{aligned}$$

Thus, the attack works.

## 9  Public Key Cryptosystems on Real Numbers

Currently used public key cryptosystems are defined over finite fields and use modular arithmetics. Their security is often based on the difficulty of solving certain number theoretic problems, such as factoring large composite integers, computing the discrete logarithm in finite multiplicative groups, deciding quadratic residuosity, computing square roots, and so on. In other words, they are designed in such a way that the cryptosystem can be broken if the presumed underlying difficult problem becomes easy to solve. At the moment, this method cannot be applied to realize chaos-based cryptosystems, since they are defined over real numbers.

Certainly two important issues must be solved in order to design a secure public key cryptosystem based on real numbers. In order to apply a reductionistic approach, some presumed difficult problem over the field of real numbers which permits implementing some one-way trapdoor function or permutation should be identified. Moreover, as the above attack in a certain way points out, the finite representation of real numbers in a computer with finite memory and the finite precision of the operations, performed by such a machine, deserve an in-depth study in order to well understand the implications in terms of security they give rise to. Paradoxically, it might also exist a good technique for implementing a secure cryptosystem over the infinite field of real numbers which turns out to be insecure for *any* finite implementation over a finite computer using a finite arithmetic precision.

Some studies dealing with the possibility of cryptographic primitives over nonclassical computational models have already been done. For example, in [7], the possibility of secret sharing schemes [4, 27] over infinite countable domains, like the set of all binary strings, was studied. It was shown that no such a scheme exist. Later on, in [8], the case of private computations over the integers was studied, and it was shown that some lower bounds that hold in the finite case do not extend to infinite domains. Recently, in [36], Cryptography over the infinite field of rational numbers, giving all parties unbounded computational power, has been considered. Under the assumption that users can sample random real numbers, and that standard field operations can be used, it turned out that secure signature and secure encryption do not exist. As well as, Diffie-Hellman key exchange, oblivious transfer, and interactive encryption.

# 10 Conclusions and Open Problems

In this paper we have analysed a public key cryptosystem based on Chebyshev polynomials. Unfortunately, even if it is efficient and based on a fascinating and elegant idea, we have shown that it is not secure, since an adversary can efficiently recover the plaintext from a given ciphertext. The proposed cryptosystem can be implemented by using any chaotic map $x_{n+1} = F_p(x_n)$ for which $F$ can be written as $F_p(x) = f(p \cdot f^{-1}(x))$, and such that $F_p(F_s(x)) = F_{p \cdot s}(x)$, i.e., it enjoys the semi-group property. Jacobian Elliptic Chebyshev Rational Maps represent another class of maps enjoying such a property. We have shown that the attack described in Section 5 can still be applied if these maps are used. Moreover, we have analysed a Diffie-Hellman like key agreement scheme based on rational maps and we have pointed out that if Jacobian Elliptic Chebyshev Rational Maps are used, then the scheme is not secure, in the sense that a passive adversary can compute the common key. Finally, we have also shown that a recently proposed authentication scheme, designed along the same lines of the public key cryptosystem, is still subject to our attack and, hence, it is not secure. The attack we have described works in *every* case in which the maps $F_p(x)$ enjoy the semi-group property, and given $x$ and $F_p(x)$, it can be efficiently computed an integer solution $p'$ to the equation $F_{p'}(x) = F_p(x)$. However, a detailed study of new implementations as well as the design and investigation of other chaos-based public key systems are interesting topics for future researches.

# 11 Acknowledgement

We would like to thank the anonymous reviewers for helpful comments and suggestions, and for pointing out references [31], [32], and [24] to us.